\documentclass[10pt]{iopart}
\usepackage{gensymb}
\usepackage{graphicx}

\usepackage{subcaption}
\usepackage{amsbsy}
\usepackage{color}
\usepackage{textcomp}
\usepackage{float}
\usepackage{sidecap}
\usepackage{siunitx}
\usepackage{amssymb}
\expandafter\let\csname equation*\endcsname\relax

\expandafter\let\csname endequation*\endcsname\relax

\usepackage{amsmath}
\usepackage{makecell}
\usepackage{upgreek}

\protected\def\PRE{\ifmmode \mathrm{PRE} \else PRE\fi}

\begin{document}

\title{I-mode pedestal relaxation events in the Alcator C-Mod and ASDEX Upgrade tokamaks}

\author{D. Silvagni${^{1}}$, J. L. Terry${^2}$, W. McCarthy${^2}$, A. E. Hubbard${^2}$, T. Eich${^1}$, M. Faitsch${^{1}}$, L. Gil${^3}$, T. Golfinopoulos${^2}$, G. Grenfell${^1}$, M. Griener${^1}$, T. Happel${^1}$, J. W. Hughes${^2}$, U. Stroth${^{1,4}}$, E. Viezzer${^5}$, the ASDEX Upgrade team$^{\mathrm{a}}$ and the EUROfusion MST1 team$^{\mathrm{b}}$}

\address{$^1$Max-Planck-Institut f{\"u}r Plasmaphysik, 85748 Garching, Germany \\ $^2$Plasma Science and Fusion Center, Massachusetts Institute of Technology, Cambridge, MA, USA \\ $^3$Instituto de Plasmas e Fus{\~a}o Nuclear, Instituto Superior T{\'e}cnico, Universidade de Lisboa, 1049-001 Lisboa, Portugal \\ $^4$Physik-Department E28, Technische Universit{\"a}t M{\"u}nchen, 85748 Garching, Germany \\ $^5$Dept. of Atomic, Molecular and Nuclear Physics, University of Seville, Avda. Reina Mercedes, 41012 Seville \\  $^{\mathrm{a}}$see author list of H. Meyer et al. 2019 Nucl. Fusion 59 112014
 \\ $^{\mathrm{b}}$see author list of B. Labit et al. 2019 Nucl. Fusion 59  086020}
 
\ead{davide.silvagni@ipp.mpg.de}
\vspace{10pt}




\begin{abstract}
In some conditions, I-mode plasmas can feature pedestal relaxation events (PREs) that transiently enhance the energy reaching the divertor target plates. To shed light into their appearance, characteristics and energy reaching the divertor targets, a comparative study between two tokamaks -- Alcator C-Mod and ASDEX Upgrade -- is carried out. It is found that PREs appear only in a subset of I-mode discharges, mainly when the plasma is close to the H-mode transition. Also, the nature of the triggering instability is discussed by comparing measurements close to the separatrix in both devices. The PRE relative energy loss from the confined region increases with decreasing pedestal top collisionality $\nu_{\mathrm{ped}}^*$. In addition, the relative electron temperature drop at the pedestal top, which is related to the conductive energy loss, rises with decreasing $\nu_{\mathrm{ped}}^*$. Finally, the peak parallel energy fluence due to the PRE measured on the divertor in both devices is compared to the model introduced in~\cite{Eich_2017} for type-I ELMs. The model is shown to provide an upper boundary for PRE energy fluence data, while a lower boundary is found by dividing the model by three. These two boundaries are used to make projections to future devices such as DEMO and ARC.

\end{abstract}

%
%
%
%
\ioptwocol

\section{Introduction}

The improved energy confinement mode, I-mode, is a promising operational regime obtained in tokamak devices. It features an enhanced energy confinement time due to the formation of a temperature pedestal at the plasma edge, while the particle confinement time and the edge density profile remain similar to those of L-mode plasmas~\cite{Ryter_1998, Whyte_2010}. In this way, high core plasma pressure and reduced impurity accumulation can be simultaneously achieved. Additionally, I-mode plasmas are free of type-I edge localized modes (ELMs), i.e. edge magnetohydrodynamic (MHD) instabilities that transiently expel energy and particles from the confined region into the scrape-off layer (SOL), leading to an increase of the heat flux onto the divertor target plates. Type-I ELMs must be avoided or mitigated in a fusion power plant, since their associated heat loads strongly reduce the lifetime of the divertor target plates~\cite{Eich_2017}. I-mode plasmas naturally lack type-I ELMs because their edge pressure profile is ideal peeling-ballooning stable~\cite{Wade_2005, Happel_2017}.
\newline However, in some particular conditions I-mode plasmas can feature small pedestal relaxation events (PREs) that transiently enhance heat fluxes on the divertor targets. These small `ELM-like' events have been first observed in the Alcator C-Mod tokamak~\cite{Whyte_2010, Walk_2014} and more recently investigated in the ASDEX Upgrade (AUG) tokamak~\cite{Silvagni_NF2020}. During I-mode PREs in AUG, about 1\,$\%$ of the total plasma energy is lost from the confined region, and the pedestal is far from the ideal peeling-ballooning stability boundary~\cite{Silvagni_NF2020}. For these reasons PREs differ from type-I ELMs, which are characterized by a pedestal close to the ideal peeling-ballooning stability boundary~\cite{Gohil_1988} and by energy losses of about 3--10$\%$ of the total plasma energy~\cite{Eich_2017}. Also, the PRE frequency of occurrence slightly increases with increasing heating power~\cite{Bonanomi_2021}, marking a difference from type-III ELMs~\cite{Zohm_1996}. In AUG, I-mode PREs were observed only just before the H-mode transition, and hence only in a restricted region of the I-mode operational space in AUG~\cite{Silvagni_NF2020}. 
\newline In order to assess the compatibility of the I-mode confinement regime with the strict requirements for component lifetimes and operation time of a fusion power plant, it is important to understand how to achieve I-mode discharges without PREs and to evaluate whether PREs are a threat for the divertor target plates, in case they appear. In this regard, multi device studies are of great help, as they provide a larger operational space under analysis and increase confidence in the predictions to a fusion reactor.
\newline In this work, I-mode PREs are investigated jointly in two tokamaks, Alcator C-Mod (C-Mod) and ASDEX Upgrade (AUG), in order to gain additional insights into the appearance of these events, the nature of the triggering instability and the energy losses reaching the divertor targets.
The paper is organized as follows. In section~\ref{sec:appearance} the appearance of PREs in both machines is described, showing typical discharges and the multi-device occurrence operational space. In section~\ref{sec:boundary} the evolution of the plasma boundary during PREs is analyzed, showing the characteristics of precursor oscillations and the subsequent pedestal relaxation. The relative plasma energy and electron temperature losses are shown in section~\ref{sec:enloss}, while section~\ref{sec:enfluence} focuses on the analysis of the energy fluence reaching the divertor target plates. Finally, in section~\ref{sec:conclusion} the main conclusions are outlined. 

\section{Appearance}
\label{sec:appearance}

\subsection{Typical discharge in C-Mod and AUG} 

\begin{figure*}[htb]
        \centerline{\includegraphics[width=1 \textwidth]{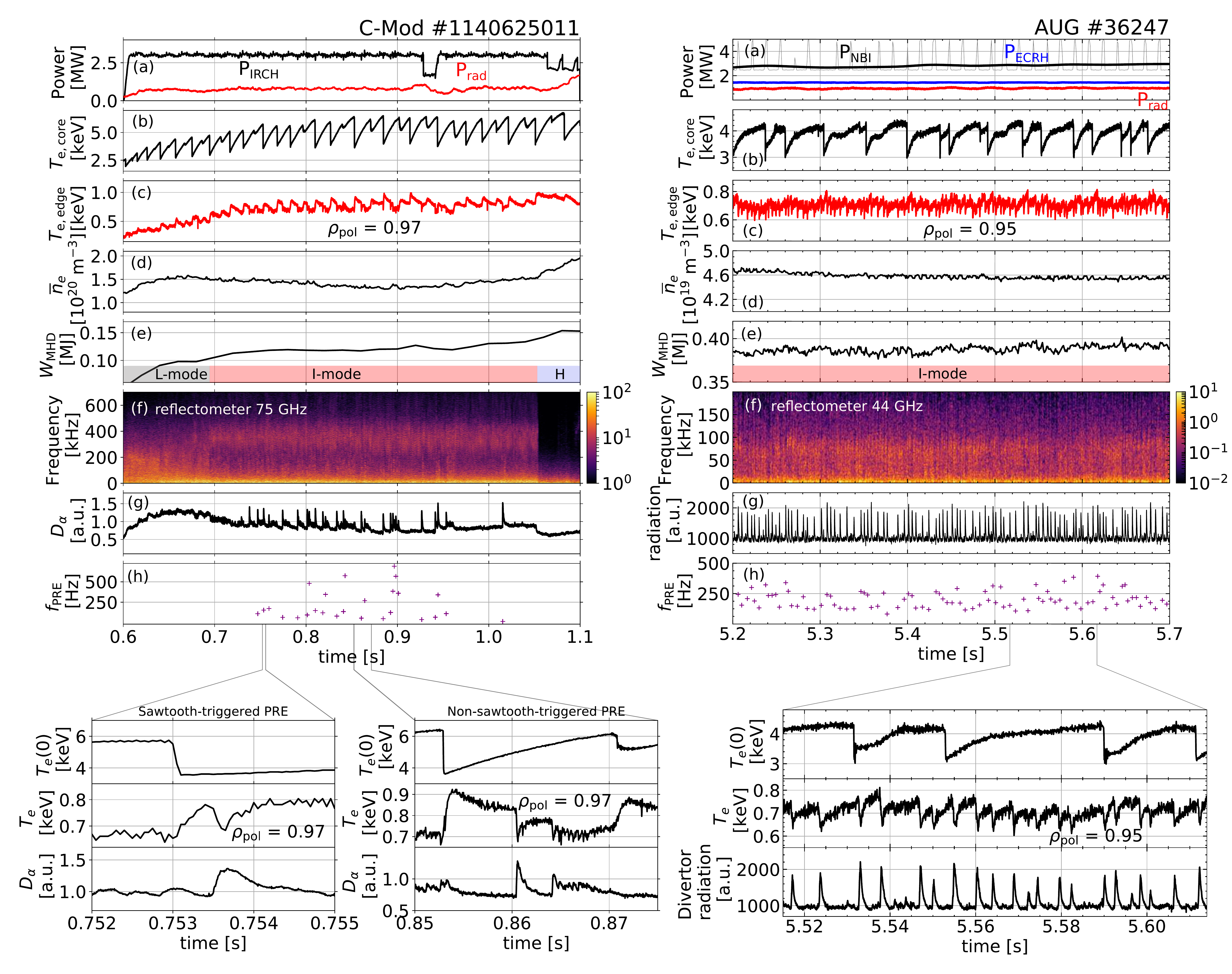}}
        \caption[]{Typical I-mode discharge with PREs in C-Mod (left) and AUG (right). (a) External heating power and radiation power (red). Electron temperature at the plasma core (b) and at the plasma edge (c) measured by ECE. (d) Line-averaged electron density measured by interferometry. (e) Plasma stored energy $W_{\mathrm{MHD}}$. (f) Spectrogram of the reflectometry signal caused by edge density fluctuations. (g) PRE monitor: for C-Mod the edge $D_{\alpha}$ light was used, while for AUG the divertor radiation measured by a diode bolometer was employed. (f) PRE frequency of occurrence. Bottom panels show magnifications of the core and edge electron temperature and of the PRE monitor in different time windows.}
        \label{overview}
\end{figure*}

Figure~\ref{overview} shows a typical I-mode discharge with PREs in C-Mod (left) and AUG (right). Both plasmas are with the ion $\nabla B$ drift pointing away from the active magnetic X-point, i.e. in the so-called unfavorable configuration in terms of H-mode access. The C-Mod shot is in lower single null (LSN) configuration at $|B_{\mathrm{t}}|=5.8$\,T and $|I_{\mathrm{p}}|=1.1$\,MA, while the AUG discharge is in upper single null (USN) configuration at $|B_{\mathrm{t}}|=2.5$\,T and $|I_{\mathrm{p}}|=1.0$\,MA. In C-Mod the plasma is heated with constant ion cyclotron resonance heating (ICRH). At the beginning of the discharge the plasma is in L-mode and later, around $t=0.68$\,s, it enters I-mode. The L-I transition is clearly visible from the rise of the edge electron temperature measured by electron cyclotron emission (ECE)~\cite{Chatterjee_2001} at $\rho_{\mathrm{pol}} = 0.97$, in panel (c), and from the rise of the plasma stored energy $W_{\mathrm{MHD}}$ evaluated from the reconstructed MHD equilibrium, in panel (e). For clarity, the definition of $\rho_{\mathrm{pol}}$, the radial coordinate used in this work, is
\begin{equation}
   \rho_{\mathrm{pol}} = \sqrt{\frac{\Psi - \Psi_{\mathrm{axis}}}{\Psi_{\mathrm{sep}} - \Psi}},
  \label{eq:rho_pol}
\end{equation}
where $\Psi$ is the poloidal magnetic flux, $\Psi_{\mathrm{axis}}$ is the flux at the magnetic axis and $\Psi_{\mathrm{sep}}$ is the flux at the separatrix.
During the L-I transition the line-averaged electron density measured by interferometry~\cite{Irby_1988} (panel (d)) stays constant, while the typical I-mode footprint - the weakly coherent mode (WCM)~\cite{Hubbard_2011, Manz_2015, Feng_2019} - appears in the spectrogram of the reflectometry signal caused by edge density fluctuations~\cite{Lin_1999} (panel (f)). During the I-mode phase some transient events - identified as I-mode PREs - are visible from the edge $D_{\alpha}$ recycling light plotted in panel (g). They have a rather intermittent occurrence frequency ranging between $\sim$\,50--600\,Hz, see panel (h). As shown in the bottom magnifications, during a PRE the edge electron temperature drops. The plasma expelled from the confined region enters the SOL, causing an increase in the edge $D_{\alpha}$ recycling light. In C-Mod, the heat pulse perturbation caused by core sawtooth instabilities can directly trigger a PRE, as shown in the bottom left magnification. Nonetheless, PREs can occur also in between two consecutive sawtooth instabilities. An example of that is shown in the central bottom magnification (note the different time scales). For this reason, the PRE frequency of occurrence can vary widely in C-Mod discharges: when PREs are only sawtooth-triggered, their frequency is regular and follows the sawtooth frequency; on the other hand, when PREs are not sawtooth-triggered, their frequency becomes more irregular. Also note that, for the same plasma conditions, sawtooth-triggered and non-sawtooth-triggered PREs exhibit a similar electron temperature drop around the pedestal top position.
\newline The right panels of Fig.~\ref{overview} show a typical AUG I-mode discharge with PREs.
The plasma is heated with constant electron cyclotron resonance heating (ECRH) and neutral beam injection (NBI), the latter being feedback controlled on the value of the $\beta_{\mathrm{pol}}$, similarly to~\cite{Happel_2019, Silvagni_NF2020}. The $\beta_{\mathrm{pol}}$ is defined as $ 2 \mu_0 \overline{p}/ \overline{B}_{\mathrm{pol}}^2$, where $\overline{p}$ is the average plasma pressure and $\overline{B}_{\mathrm{pol}}$ is the average poloidal magnetic field strength. The plasma is in I-mode for the whole time window shown here, which can be seen by the presence of the WCM in the spectrogram of the reflectometry signal~\cite{Silva_1996} probing at $\rho_{\mathrm{pol}} \approx 0.98$ (panel (f)). During this I-mode phase, the core and edge electron temperature are rather constant, as the plasma was feedback controlled to stay around a fixed $\beta_{\mathrm{pol}}$ value. Since PREs in AUG have been observed only before the H-mode transition~\cite{Silvagni_NF2020}, the target $\beta_{\mathrm{pol}}$ value was chosen to be just slightly below the one at which an I-H transition would have been triggered. For this reason, during this discharge PREs are always present, as can be seen in panel (g) from the divertor radiation measured by a diode bolometer. Their frequency of occurrence ranges between 100 and 400 Hz and it is more regular than that observed in C-Mod. The magnification on the bottom shows that in AUG the sawtooth cycle is not strongly impacting the occurrence of PREs, contrary to the C-Mod case.

\subsection{Multi-device operational space} 

To study the occurrence of I-mode PREs, a database of I-mode discharges from both C-Mod and AUG has been assembled. The main plasma parameters of the database are shown in table~\ref{database_all}. For C-Mod, 375 I-mode discharges have been included, of which 79 have a phase with PREs, whereas regarding AUG 109 I-mode discharges have been selected, of which 26 with PREs. These figures highlight that PREs usually appear in a restricted subset of I-mode discharges. In the analyzed database, only about 20\,$\%$ of I-mode discharges have a phase with PREs.
\newline With regard to AUG, the database used in Ref.~\cite{Silvagni_NF2020} 
\begin{table}[htb]
    {\footnotesize \centerline{
    \begin{tabular}{l|cc}
        \hline\hline
         &  \makecell{C-Mod } & \makecell{AUG } \\
        \hline
        $\#$ Discharges & 375 & 109 \\
        $W_{\mathrm{MHD}}$ (MJ) & 0.03--0.23 & 0.18--0.45 \\
        $V (\mathrm{m}^{-3})$ & 0.8--1.0 & 11--13 \\
        $\overline{n}_e$ ($10^{19}$ m$^{-3}$) & 7.5--23 & 2.4--7.3  \\
        $p_{e \mathrm{, ped}}$ (kPa) & 1.3--24 & 1.3--4.3 \\
        $T_{e \mathrm{, ped}}$ (keV) & 0.2--1.2 & 0.2--1.0 \\
        $n_{e \mathrm{, ped}}$ ($10^{19}$ m$^{-3}$) & 3.0--16 & 1.5--6.0 \\
        $q_{95}$ & 2.8--6.0 & 3.6--8.2 \\
        $I_{\mathrm{p}}$ (MA) & 0.6--1.7 & 0.6--1.0 \\
        $B_{\mathrm{t}}$ (T) & 2.8--8.0 & 1.8--3.2 \\
        $\delta$ & 0.4--0.9 & 0.2--0.3 \\
        $\beta_{\mathrm{pol}}$ & 0.3--1.4 & 0.3--1.4 \\
        $\nu_{\mathrm{ped}}^*$ & 0.1--2.9 & 0.1--5.1 \\
        $f_{\mathrm{GR,ped}}$ & 0.04--0.20 & 0.15--0.55 \\
        \hline\hline
    \end{tabular}}}
    \caption[]{Parameter range of the I-mode AUG and C-Mod discharges analyzed. The plasma volume within the separatrix is denoted by $V$ and the plasma triangularity is $\delta = (\delta_{u} + \delta_{l})/2$, where $\delta_u$ and $\delta_l$ are the upper and lower triangularity, respectively. The pedestal top collisionality $\nu_{\mathrm{ped}}^*$ is defined in Eq.~\ref{eq:pedestal_coll}, while the pedestal Greenwald fraction is $f_{\mathrm{GR,ped}} = n_{e \mathrm{, ped}} / n_{\mathrm{GR}}$, where the Greenwald density is $n_{\mathrm{GR}} = I_p / (\pi a^2 )$.}
    \label{database_all}
\end{table}
has been re-examined and enlarged with I-mode discharges at different toroidal magnetic field strengths and plasma currents. Electron temperature and density values have been obtained through integrated data analysis (IDA)~\cite{Fischer_2010}, which combines measurements from the edge and core Thomson scattering system~\cite{Kurzan2011}, electron cyclotron emission radiometers~\cite{Denk_2018} and DCN interferometry~\cite{Mlynek_2010}. 
\newline Concerning C-Mod, the databases used in Refs.~\cite{Whyte_2010} and~\cite{Hubbard_2016} have been analyzed and extended by including more recent I-mode discharges of the 2016 campaign~\cite{Hubbard_2017} and by identifying time windows with and without PREs. Edge electron temperature and density profiles have been obtained by fitting Thomson scattering~\cite{Hughes_2003} and ECE data to the modified hyperbolic tangent function~\cite{Groebner_2001}, similarly to~\cite{Whyte_2010}. In order to guarantee enough data points for a reliable fit, time windows larger than 20\,ms have been used. For this reason, the profiles obtained should be regarded as an average over one (or more) sawtooth cycles.
\begin{figure}[htb]
        \centerline{\includegraphics[width=0.5 \textwidth]{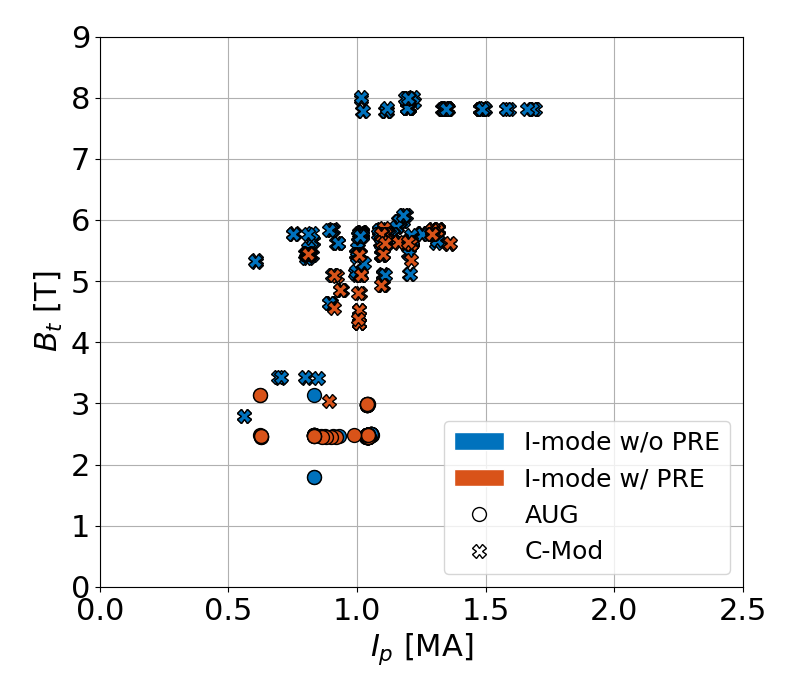}}
        \caption[]{Plasma current $I_{\mathrm{p}}$ and toroidal magnetic field at the magnetic axis $B_{\mathrm{t}}$ of the analyzed discharges in C-Mod (crosses) and AUG (circles). I-mode discharges with and without PREs are marked in orange and blue, respectively. Note that at $B_{\mathrm{t}}=8$\,T no PREs have been observed in C-Mod.}
        \label{Ip_Bt}
\end{figure}
\newline Figure~\ref{Ip_Bt} shows the plasma current $I_{\mathrm{p}}$ and toroidal magnetic field $B_{\mathrm{t}}$ values of the discharges under analysis for both devices. The two machines span a large range of $B_{\mathrm{t}} \in [1.8, 8]$\,T and $I_{\mathrm{p}} \in [0.5, 1.7]$\,MA. Interestingly, at $B_{\mathrm{t}} \approx 8$\,T no PREs have been observed in any of the inspected I-mode discharges in C-Mod. It should be noted that, in unfavorable configuration, no H-mode transitions were found in any of the 8\,T C-Mod discharges ~\cite{Hubbard_2017}. This is because the available external heating power of 5\,MW was below the required I-H power threshold, which is known to increase roughly linearly with $B_{\mathrm{t}}$~\cite{Martin_2008}. This is a first indirect suggestion that in C-Mod, like in AUG, PREs are more likely to exist near the I-H transition boundary. To show this more clearly, a proxy for the H-mode transition which could be used across different $I_{\mathrm{p}}$ and $B_{\mathrm{t}}$ needs to be introduced.
\newline It is widely accepted that the formation of the edge transport barrier, which is the characteristic feature of H-mode, arises from turbulence suppression due to an increase of the $E \times B$ velocity shear at the plasma edge. In support of this theory, an empirical threshold for the minimum $v_{E \times B}$ necessary for the H-mode onset has been recently found in AUG~\cite{Cavedon_2020}. Therefore, one could use the $v_{E \times B}$ minimum at the plasma edge as a proxy for the H-mode transition:
\begin{equation}
   v_{E \times B, \mathrm{min}} = \frac{E_r}{B},
  \label{eq:vmin}
\end{equation}
where $E_r$ is the radial electric field at the location of the $v_{E \times B}$ minimum. Assuming that the main contribution to the edge $E_r$ comes from the neoclassical radial electric field~\cite{Sauter_2011, Viezzer_2013, McDermott_2009}, which is mainly set by the diamagnetic term~\cite{Hinton_1976}, one can approximate: 
\begin{equation}
   E_{r} \approx \frac{\nabla p_i }{q_i n_i},
  \label{eq:Er}
\end{equation}
where $p_i$, $n_i$ and $q_i$ are the main ion pressure, density and charge, respectively. Assuming $T_i=T_e$ and $n_e=n_i$, and considering that the edge pressure gradient is proportional to the pedestal top value if the pedestal width and separatrix values do not vary significantly, Eq.~\ref{eq:Er} can be further simplified as follows:
\begin{equation}
   E_{r} \propto \frac{ p_{e, \mathrm{ped}} }{ n_{e, \mathrm{E_r min}}} \approx \frac{ p_{e, \mathrm{ped}} }{ n_{e, \mathrm{sep}}},
  \label{eq:Er_simply}
\end{equation}
where $ p_{e, \mathrm{ped}} $ is the electron pressure at the pedestal top and $n_{e, \mathrm{E_r min}}$ is the electron density at the location of the $E_r$ minimum, which here it is assumed to be similar to the electron density at the separatrix $n_{e, \mathrm{sep}}$.
Therefore, our proxy for the H-mode transition, the $v_{E \times B}$ minimum, can be approximated as: 
\begin{equation}
   v_{E \times B, \mathrm{min}} \propto  \frac{ p_{e, \mathrm{ped}} }{ B_t n_{e, \mathrm{sep}}}.
  \label{eq:vmin_simply}
\end{equation}
In other words, the H-mode transition happens above a certain value of $p_{e, \mathrm{ped}}$ normalized to the toroidal magnetic field and the electron density at the separatrix.
\begin{figure}[htb]
        \centerline{\includegraphics[width=0.5 \textwidth]{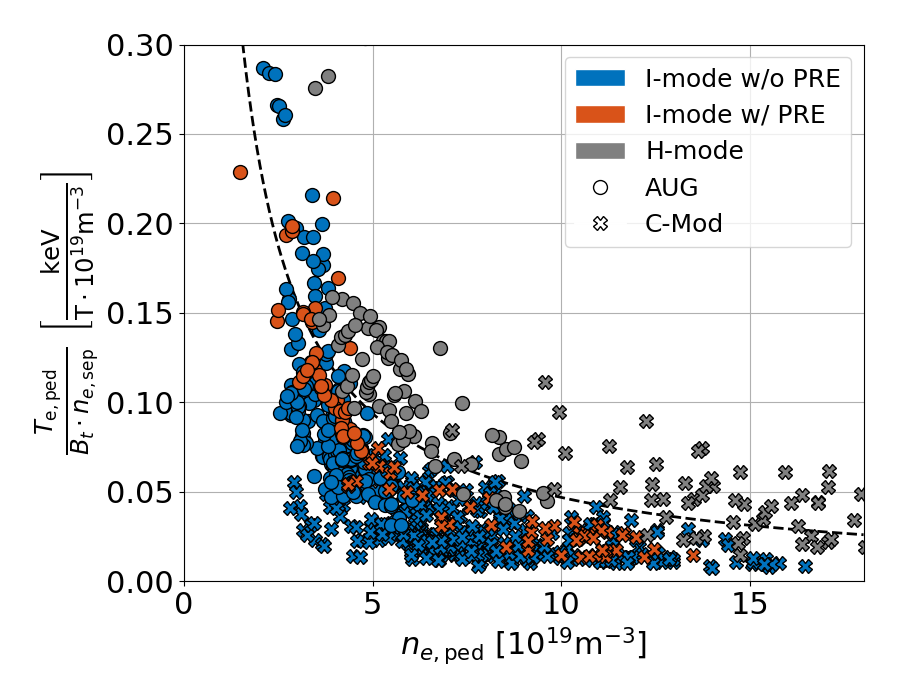}}
        \caption[]{Electron temperature at the pedestal top $T_{e, \mathrm{ped}}$ normalized to the toroidal magnetic field $B_{\mathrm{t}}$ and to the separatrix electron density $n_{e, \mathrm{sep}}$ against the electron density at the pedestal top $n_{e, \mathrm{ped}}$ for C-Mod (crosses) and AUG (circles) discharges. I-mode discharges with and without PREs are marked in orange and blue, respectively, while H-mode discharges are colored in gray. I-mode PREs tend to appear close to the H-mode transition. The dashed line represents $p_{e, \mathrm{ped}} / (B_t n_{e, \mathrm{sep}} )= 0.75$\,\,kPa/(T\,$\cdot$\,10$^{19}$\,m$^{-3}$).}
        \label{Te95norm_ne95}
\end{figure}
Figure~\ref{Te95norm_ne95} shows the $T_{e, \mathrm{ped}}$--$n_{e, \mathrm{ped}}$ operational space for both machines, where the electron temperature has been normalized to the $B_t n_{e, \mathrm{sep}}$ factor. Therefore, $ y= k /x$ lines in Fig.~\ref{Te95norm_ne95} (where $k$ is a constant) are an approximation of the H-mode transition threshold defined in Eq.~\ref{eq:vmin_simply}. The equation 
\begin{equation}
   \frac{ p_{e, \mathrm{ped}} \, [\mathrm{kPa}] }{ B_t \,[\mathrm{T}] \cdot n_{e, \mathrm{sep}} \,[10^{19} \, \mathrm{m}^{-3}]} = 0.75
  \label{eq:H_thresh}
\end{equation}
is plotted as a dashed line in Fig.~\ref{Te95norm_ne95}.
Overall, I-mode with PREs (orange) occur close to H-mode points for both devices, generalizing the result found in~\cite{Silvagni_NF2020} which used AUG data only, i.e. that I-mode PREs appear when the I-mode plasma is close to the H-mode transition. This also explains why large sawtooth instabilities can trigger a PRE: indeed, the local edge temperature increase induced by the heat pulse perturbation of a sawtooth instability can bring the plasma closer to the H-mode transition or even trigger the transition itself, see e.g.~\cite{Cziegler_2013}. 
Also note that the scatter in the C-Mod data with PREs might be due to the data-averaging process over multiple sawtooth cycles, which has the effect of lowering the edge temperature with respect to the one before the onset of a sawtooth-triggered PRE, see e.g. Fig.~\ref{overview}. 
\newline The implication of this finding is that PREs can be avoided if the I-mode discharge is far enough from the H-mode transition and if the heat pulse caused by a sawtooth instability does not bring the plasma close to the H-mode transition. Moreover, the appearance of PREs could be used to monitor the proximity to H-mode and to avoid the plasma entering an undesired ELMy H-mode.

\subsection{Additional considerations on the I-H transition}

A previous joint AUG/C-Mod study on the L-H transition in the favorable configuration showed that a critical ion heat flux per particle is needed to access H-mode, and that this critical value scales with the magnetic field~\cite{Schmidtmayr_2018}. This finding is consistent with a paradigm of the H-mode transition occurring when a critical value of edge $E \times B$ velocity shear $ \sim E_r /B$ is approached. Moreover, recent measurements of the $E \times B$ velocity minimum at the H-mode transition in AUG showed that a critical value of about 6.7\,km/s is needed to access H-mode at different magnetic fields and with different isotope fueling~\cite{Cavedon_2020}. However, these results were found in the favorable configuration and it is still unclear if such critical values are also needed to access H-mode in the unfavorable configuration. Previous studies on the I-mode power window of existence have shown that the I-H power threshold strongly depends on $B_t$~\cite{Hubbard_2016, Hubbard_2017}, similarly to the L-H power threshold in the favorable configuration~\cite{Martin_2008}. Therefore, this strong $B_t$ dependence might suggest that the edge $E \times B$ velocity shear could play an important role for the H-mode access also in the unfavorable configuration.
\newline It is interesting to re-examine Fig.~\ref{Te95norm_ne95} in light of the above discussion. Figure~\ref{Te95norm_ne95} shows that the H-mode proxy defined by Eq.~\ref{eq:H_thresh} divides well I-mode and H-mode data points for $n_{e \mathrm{, ped}} > 4 \times  10^{19}$\,m$^{-3}$ in both machines for a wide range of magnetic fields and plasma currents. The discrepancy at lower densities for the AUG data might be explained by possible differences between $T_i$ and $T_e$ profiles, which in this analysis are neglected. Indeed, to obtain low density discharges in AUG, only ECRH external heating is typically applied. In these conditions, $T_e > T_i$ at the pedestal top~\cite{Ryter_2014} and this will lead to $p_{e, \mathrm{ped}} / (B_t n_{e, \mathrm{sep}} ) > p_{i, \mathrm{ped}} / (B_t n_{i, \mathrm{sep}} ) $, with the right hand side being the H-mode transition proxy when the approximation $T_e=T_i$ is not introduced. This might explain why $p_{e, \mathrm{ped}} / (B_t n_{e, \mathrm{sep}} ) $ exceeds the empirical threshold defined by Eq.~\ref{eq:H_thresh} for low density AUG discharges.
\newline The overall good division of I-mode and H-mode points given by Eq.~\ref{eq:H_thresh} might suggest that also in the unfavorable configuration the minimum of the $E \times B$ velocity could play an important role for the H-mode access. Indeed, in AUG the $E_r$ minimum at the I-H transition is around 15--20\,kV/m at $B_t = 2.5$\,T~\cite{Happel_2017}, which gives $v_{E \times B \mathrm{, min}} = 6-8$\,km/s. These values of $v_{E \times B \mathrm{, min}}$ are very similar to the empirical threshold $v_{E \times B \mathrm{, min}}= 6.7$\,km/s found in favorable configuration at AUG~\cite{Cavedon_2020}.


\section{Plasma edge evolution during PREs}
~\label{sec:boundary}
\begin{figure*}[htb]
        \centerline{\includegraphics[width=1 \textwidth]{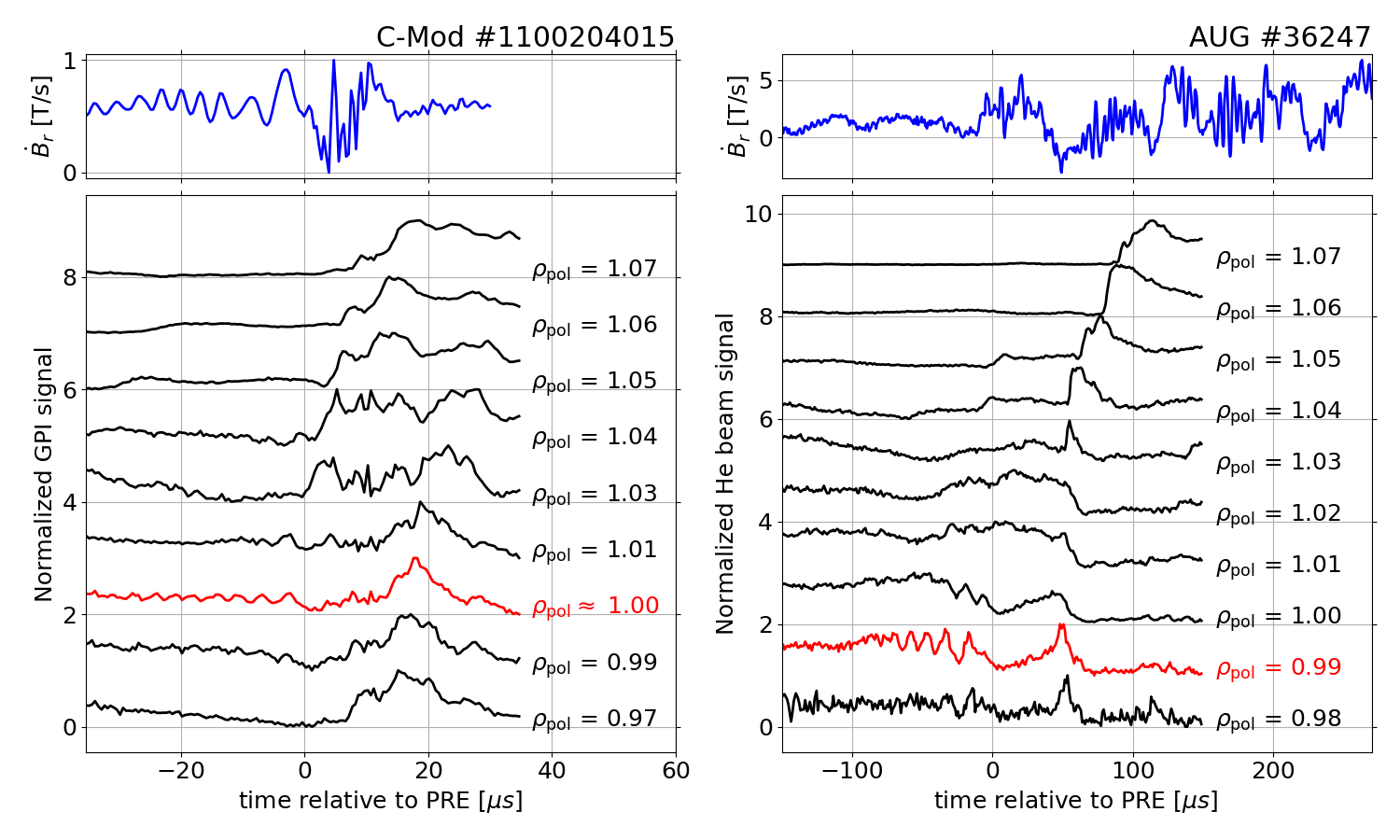}}
        \caption[]{Top: Time derivative of the radial component of the magnetic field measured by a pick-up coil located at the low field side outer midplane in C-Mod (left) and AUG (right) during a PRE. Bottom: Normalized gas puff imaging (left) and helium 587.6\,nm line intensity (right) signal at different $\rho_{\mathrm{pol}}$ during a PRE. An artificial offset has been added to the signals to improve the visibility. The onset time of the PRE is $t=1.103$\,s for C-Mod and $t=5.745$\,s for AUG. A growing precursor oscillation is visible in the edge confined region close to the separatrix (channels displayed in red).}
        \label{precursor}
\end{figure*}

\noindent To shed light into the nature of the triggering instability of a PRE, it is of interest to study how the edge plasma evolves before the PRE onset in both machines. For this purpose, 2D imaging at the outboard midplane edge region during a PRE crash using the Gas-Puff Imaging (GPI) diagnostic~\cite{Cziegler_2010} is analyzed. Its high temporal and spatial resolution give additional insights into the electron density and temperature perturbation during a PRE. In C-Mod, the GPI system puffs helium or deuterium gas locally in the plasma edge. Atomic line radiation, excited by the interaction with the ``background'' plasma electrons, serves as a proxy for the fluctuations/perturbations occurring in the ``background'' plasma. The visible line radiation is then measured with a 2\,MHz acquisition frequency by high sensitivity avalanche photodiodes (APD) that have a 2D view on the low field side (LFS) plasma edge region (9x10 poloidal-radial grid). In AUG, the equivalent of the GPI signal is provided by the thermal helium beam diagnostics~\cite{Griener_2017}, which puffs thermal helium into the plasma edge and collects the light emitted by the excited triplet and singlet helium states, which is also a function of the electron density and temperature. The system has a temporal resolution of 900\,kHz and a 2D view in the LFS plasma edge region with 32 lines of sight~\cite{Griener_2018}.
\newline The bottom panels of Fig.~\ref{precursor} show the time evolution of the GPI signals measured at different radial positions and same poloidal position in the outer midplane of both C-Mod (left) and AUG (right). Each radial position is indicated with the normalized radial coordinate $\rho_{\mathrm{pol}}$ defined in Eq.~\ref{eq:rho_pol} and the line intensity of each channel is normalized to its maximum value. Each channel has been plotted with an artificial offset to improve the visibility of signals.
In C-Mod, the separatrix position has been shifted with respect to the one given by the equilibrium reconstruction code so that the separatrix occurs where the plasma poloidal velocity changes sign. Indeed, the plasma flows in the electron diamagnetic drift direction (EDD) in the edge confined region, while in the SOL it changes direction, flowing in the ion diamagnetic drift direction (IDD)~\cite{Cziegler_2010, Theiler_2017, Happel_2017, Conway_2011}. In AUG, the separatrix position is given by the equilibrium reconstruction code. After the onset of the PRE, structures propagating radially outwards, the so-called filaments, develop in the edge-SOL region. In AUG, filaments during PREs are propagating with a radial velocity of about 0.5\,km/s~\cite{Silvagni_NF2020}, while in C-Mod their radial velocity is around 3\,km/s. 
\newline Before the onset of the PRE, a growing oscillating precursor is measured in the edge confined region very close to the separatrix in both devices (GPI channels highlighted in red). The magnetic signature of this precursor has not been detected in AUG pick-up coils yet, and it is typically not detected in C-Mod either. However, when the distance between the separatrix and the outboard limiter, which is where the magnetic pick-up coils are located, was reduced to 6.5\, mm, the precursor oscillation was detected in C-Mod. An example of such discharge is shown in Fig.~\ref{precursor}. Please note that the PRE shown here is not sawtooth-triggered and, hence, the precursor is not connected to a core sawtooth instability. This demonstrates that the growing instability causing the PRE is electromagnetic. It is therefore highly probable that the reason the magnetic precursor is not detected in AUG and in the typical C-Mod discharges is because the magnetic pick-up coils are not close enough to the separatrix to measure the perturbation. Indeed, typical outer gaps in AUG and C-Mod I-mode discharges are around 5\,cm and 1.5\,cm, respectively, i.e. much larger than 6.5\,mm. Also, it should be noted that in C-Mod, when the outer gap was 9.5\,mm, the precursor was not detected in the pick-up coils. No discharges with PREs and with the outer gap between 6.5 and 9.5 mm are present in the analyzed database. Therefore, in C-Mod the minimum distance between the separatrix and the outboard limiter necessary to detect the precursor in the pick-up coils is in the range 6.5--9.5\,mm. This is consistent with a short radial wavelength of the precursor oscillation.
\newline The detection of the precursor with the C-Mod pick-up coils allowed investigations of the toroidal mode number by comparing the phase difference between signals measured at different toroidal locations. The precursor has a toroidal mode number of $n=$\,10--20 and is rotating counter-clockwise, i.e. counter-$I_{\mathrm{p}}$ and counter-$B_{\mathrm{t}}$. The toroidal mode numbers found in C-Mod are very similar to the `quasi mode number' $n_{\mathrm{QMN}}$~\cite{Eich_2005} evaluated with infrared thermography in AUG during PREs~\cite{Silvagni_NF2020}, i.e. $n_{\mathrm{QMN}} = 18$, and to the WCM toroidal mode number measured in C-Mod, i.e. $n_{\mathrm{WCM}} \simeq 20$~\cite{Liu_2016}.
\newline In addition, the precursor propagates 
\begin{figure*}[htb]
        \centerline{\includegraphics[width=1 \textwidth]{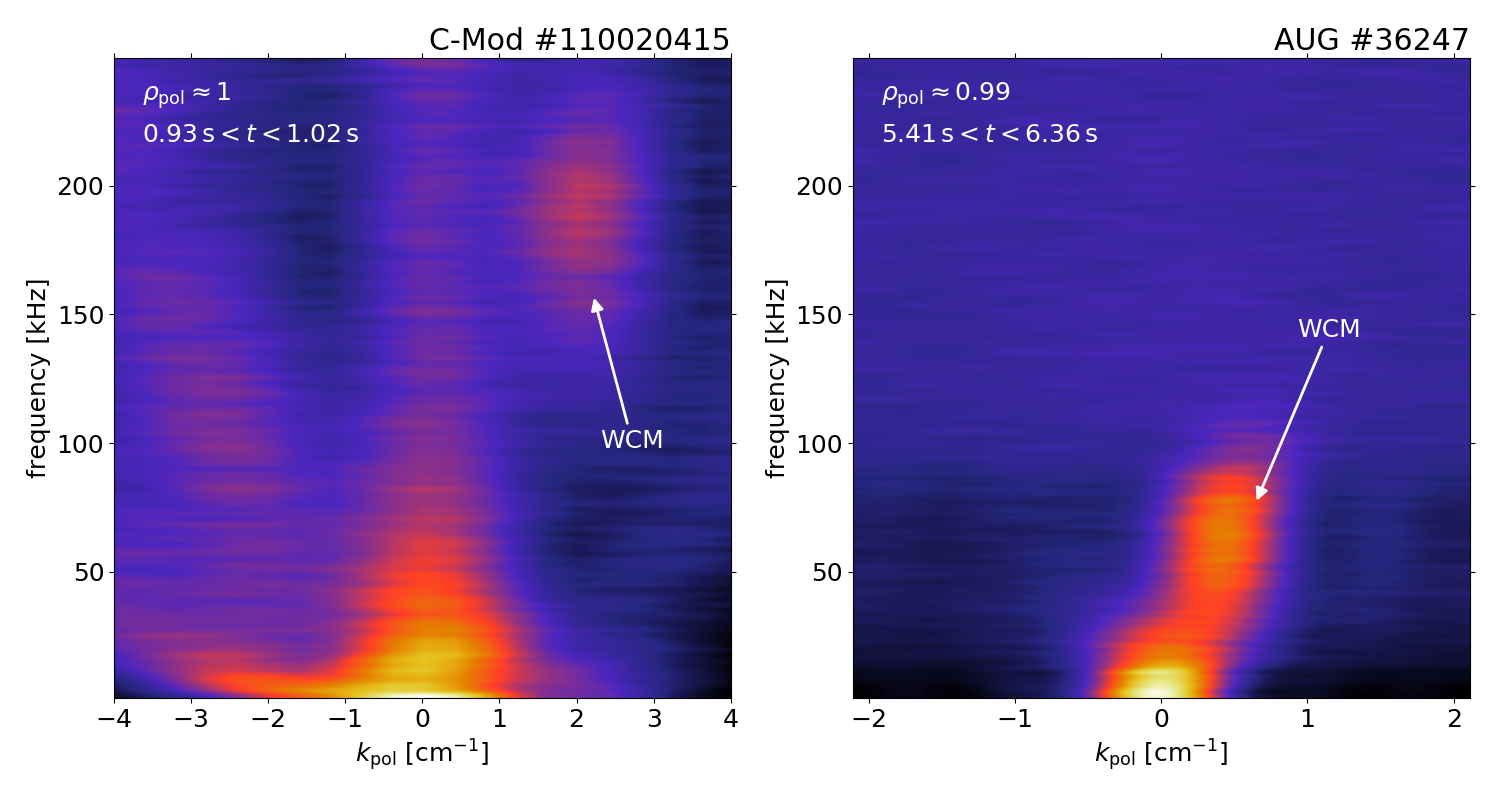}}
        \caption[]{Conditional wavenumber-frequency spectra from the edge region of I-mode plasmas in C-Mod (left) and AUG (right). The spectra are obtained from GPI (left) and the helium 587.6\,nm line intensity (right) signals. The WCM peaks at different frequency and wavenumbers in the two machines.}
        \label{kfspectrum}
\end{figure*}
poloidally in the EDD direction in both devices, but with different frequencies, i.e. $f_{\mathrm{prec}} \simeq 300$\,kHz in C-Mod and $f_{\mathrm{prec}} \simeq 75$\,kHz in AUG. 
To compare these frequencies to the WCM frequency, the conditional wavenumber-frequency spectra of the GPI signal in C-Mod and of the thermal helium beam signal in AUG are shown in Fig.~\ref{kfspectrum}. These spectra have been calculated using the vertical/poloidal channels at the radial location where the precursor occurs (red lines in Fig.~\ref{precursor}). The conditional $k$--$f$ spectrum $S(k_{\mathrm{pol}}|f)$ is the $k$--$f$ spectrum $S(k_{\mathrm{pol}, f})$ normalized to the spectrum at each frequency band $S(f) = \sum_{k_{\mathrm{pol}}} S(k_{\mathrm{pol}, f})$. This normalization is helpful to highlight trends in the high frequency--high wavenumber domain, since turbulence spectra fall off rather quickly with $f$ and $k$. In the notation used here, positive $k_{\mathrm{pol}}$ denotes upwards propagation, i.e. a flow in the EDD direction in these USN discharges in the unfavorable configuration. 
The WCM is propagating in both devices in the EDD direction in the laboratory frame (as the precursor does), and it shows different characteristics in the two machines: in C-Mod it peaks around $f\approx 200$\,kHz and $k_{\mathrm{pol}}\approx 2$\,cm$^{-1}$, while in AUG it appears around $f\approx 70$\,kHz and $k_{\mathrm{pol}}\approx 0.5$\,cm$^{-1}$. Both frequency and $k_{\mathrm{pol}}$ values of the WCM are in the range of typical C-Mod and AUG values~\cite{Cziegler_2013, Manz_2015}. Therefore, in AUG the PRE precursor oscillates with a frequency very similar to the one of the WCM, whereas in C-Mod it oscillates with a frequency roughly 1.5$\times$ larger than that of the WCM.  

\subsection{Discussion on the PRE-triggering instability}

According to the simulations in Ref.~\cite{Manz_2021}, a PRE can be triggered when the plasma beta around the separatrix becomes large enough to induce radial magnetic incoherent fluctuations. These perturbations disturb the parallel electron dynamic which is central for the formation of the WCM in those simulations~\cite{Manz_2020} and which is stabilizing for interchange effects. As a consequence, the WCM becomes interchange unstable and the associated enhanced transport relaxes edge gradients causing a PRE. Therefore, according to the simulation, a PRE develops from the WCM which becomes interchange unstable at large plasma beta. This interpretation is consistent with several experimental observations presented here. First of all, in AUG after the disappearance of the precursor, radial magnetic field fluctuations becomes more incoherent, see Fig.~\ref{precursor}, similarly to what has been found in the simulation~\cite{Manz_2021}. No clear statement can be made on the C-Mod cases, as the GPI diagnostics and magnetic pick-up coils are not time synchronized. Moreover, the precursor oscillation and the WCM share similar toroidal mode numbers and, in AUG, also a similar frequency of oscillation. In C-Mod, the frequency of the precursor is higher than that of the WCM. However, it should be noted that the C-Mod $k$--$f$ spectrum in Fig.~\ref{kfspectrum} has been calculated in an I-mode time window without PREs. Therefore, the discrepancy between the precursor frequency and the WCM frequency might be explained by a transient increase of the radial electric field, which is a typical feature happening when the plasma gets closer to the H-mode transition. This would transiently increase the background $E \times B$ flow and thus the frequency of the mode.
Concerning the appearance of PREs, a local increase of the plasma beta around the separatrix would be expected when the plasma approaches H-mode. Unfortunately, the analyzed database does not allow precise evaluation of the plasma beta around the separatrix. Hence, this hypothesis cannot be experimentally confirmed with this dataset.

\section{Relative energy losses from the confined region}
\label{sec:enloss}

\begin{figure}[htb]
        \centerline{\includegraphics[width=0.5 \textwidth]{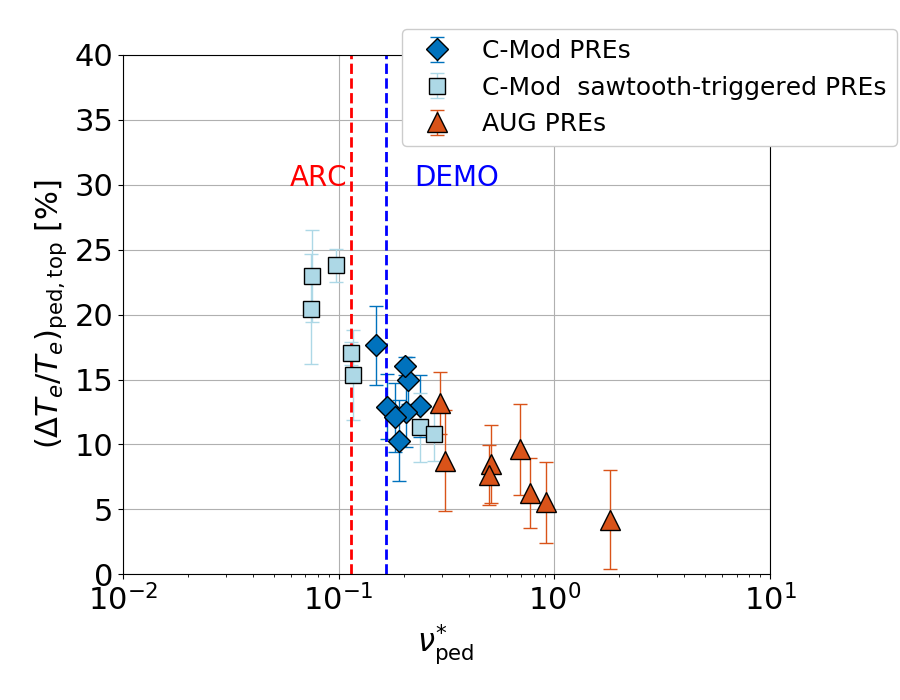}}
        \caption[]{Relative loss of the pedestal top electron temperature during PREs against the pedestal top collisionality in C-Mod and AUG. C-Mod PREs are broken down into sawtooth-triggered (squares) and non-sawtooth-triggered (diamonds) PREs. The relative electron temperature drop (proportional to the conductive losses) is well correlated with the pedestal top collisionality.}
        \label{Tedrop}
\end{figure}

During a PRE the edge pressure profile relaxes and, hence, part of the plasma energy exits the confined region and enters the SOL. Ultimately, a fraction of this energy reaches the divertor target plates, enhancing transiently the heat fluxes. Therefore, it is of interest to study how the PRE energy loss from the plasma confined region changes with plasma parameters. 
The energy loss $\Delta W$ during a transient event is defined as:
\begin{equation}
    \Delta W = W(t_0) -  W(t_1) = \frac{3}{2} \int \Delta p \,\mathrm{d}V,
    \label{eq:enloss}
\end{equation}
where $t_0$ and $t_1$ are the time instants before and after the transient event, respectively, $V$ is the volume and $p= p_i + p_e$ is the sum of the ion and electron pressure. To allow cross-machine comparisons, the energy losses are usually normalized either to the total plasma stored energy $W_{\mathrm{MHD}}$ evaluated from equilibrium reconstruction codes or to the pedestal energy, defined as $W_{\mathrm{ped}} = 3/2 \,p_{\mathrm{ped}} V_{\mathrm{plasma}}$, where $V_{\mathrm{plasma}}$ is the plasma volume within the separatrix. Previous studies on type-I ELMs highlighted an inverse correlation between the size of the relative energy losses and the pedestal top collisionality~\cite{Loarte_2003}, defined as:
\begin{equation}
    \nu^*_{\mathrm{ped}}= 6.9 \times 10^{-18} \frac{R \, q_{\mathrm{95}} \, Z_{\mathrm{eff}}  \, \mathrm{ln}\Lambda \, n_{e, \mathrm{ped}}} {\epsilon^{3/2} \, T_{e, \mathrm{ped}}^2},
    \label{eq:pedestal_coll}
\end{equation} 
where $T_{e, \mathrm{ped}}$ and $n_{e, \mathrm{ped}}$ are the pedestal top electron temperature and density in eV and m$^{-3}$, respectively, $R$ is the major radius, $q_{95}$ is the safety factor at $\rho_{\mathrm{pol}}=0.95$, $Z_{\mathrm{eff}}$ is the effective ion charge, $\epsilon$ is the inverse aspect ratio and $\mathrm{ln}\Lambda = 31.3 - \mathrm{ln} ( \sqrt{n_{e, \mathrm{ped}}}/T_{e, \mathrm{ped}} )$ is the quantum Coulomb logarithm at the pedestal top~\cite{Honda_2013}. 
The energy loss can be further divided into two components, one - called conductive loss - which is related to the temperature loss $\Delta T$, and another one - called convective loss - which is related to the density loss $\Delta n$:
\begin{gather}
  \nonumber \Delta W  \approx  \frac{3}{2} k \left[\int n \, \Delta T \, \mathrm{d}V + \int T \, \Delta n \, \mathrm{d}V \right] = \\
  = \Delta W_{\mathrm{cond}} + \Delta W_{\mathrm{conv}}.
  \label{eq:cond_conv}
\end{gather}
The cross-term has been neglected because it is of second order. The distinction between conductive and convective losses is important because these two terms could scale differently to a fusion power plant. Indeed, it has been shown for type-I ELMs that conductive losses strongly depend on pedestal parameters (such as collisionality), while convective losses exhibit a weak dependence on pedestal parameters~\cite{Leonard_2002, Beurskens_2009}.
\begin{figure}[htb]
        \centerline{\includegraphics[width=0.5 \textwidth]{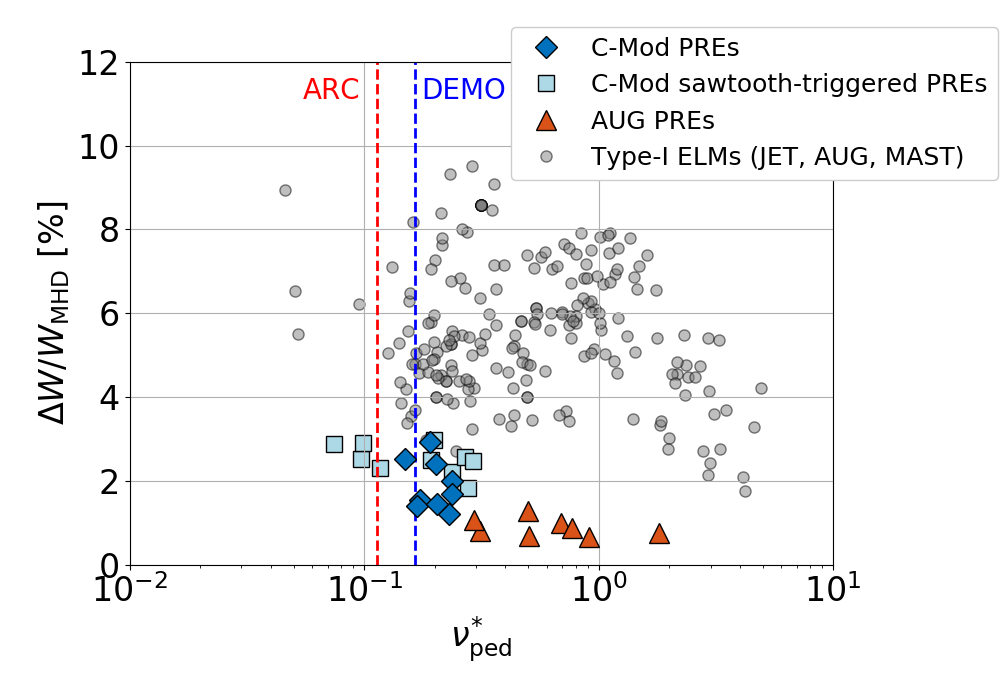}}
        \caption[]{PRE relative energy loss against the pedestal top collisionality in C-Mod and AUG (triangles). C-Mod PREs are broken down into sawtooth-triggered (squares) and non-sawtooth-triggered (diamonds) PREs. Type-I ELM data from the multi device study in Ref.~\cite{Eich_2017} are plotted for comparison. The dashed lines indicate the expected I-mode pedestal top collisionality in DEMO~\cite{Siccinio_2021, Morris_2021} and ARC~\cite{Sorbom_2015}.}
        \label{Wmhdloss}
\end{figure}
\newline In this work energy losses have been calculated directly from the change of the $W_{\mathrm{MHD}}$ signal and not from profile reconstruction. 
Figure~\ref{Wmhdloss} shows the PRE energy loss normalized to $W_{\mathrm{MHD}}$ at the PRE onset against the pedestal top collisionality in AUG and C-Mod. Data of type-I ELMs from the multidevice study in Ref.~\cite{Eich_2017} are plotted in light grey for comparison. The energy losses have been calculated making a conditional average of the $W_{\mathrm{MHD}}$ signal during several PREs in discharge phases with constant plasma parameters, similarly to~\cite{Silvagni_NF2020}.
The PRE relative energy loss increases with decreasing pedestal top collisionality, similarly to type-I ELMs. However, a clear difference from type-I ELMs is the size of the relative losses, which ranges between 0.5 and 3\,$\%$ for PREs, while it is mainly between 3 and 10\,$\%$ for type-I ELMs. 
Also, in Fig.~\ref{Wmhdloss} C-Mod data are broken down into sawtooth-triggered (squares) and non-sawtooth-triggered (diamonds) PREs. No significant difference has been found in the relative energy loss of sawtooth-triggered and non-sawtooth-triggered PREs for discharges with similar plasma parameters. The trend found in Fig.~\ref{Wmhdloss} highlights that the PRE energy loss reaches its larger values, i.e. about 2--3\,$\%$ of the total plasma energy, at low pedestal top collisionalities. These are also the values of collisionalitites expected at the I-mode pedestal top in the European DEMO~\cite{Siccinio_2021, Morris_2021} and ARC~\cite{Sorbom_2015} tokamaks.
\newline In order to evaluate how PREs conductive and convective losses scale with plasma parameters, Eq.~\ref{eq:cond_conv} is further simplified by writing the volume integrals as:
\begin{gather}
  \nonumber \int n \, \Delta T \, \mathrm{d}V  = \alpha \, n_{\mathrm{ped}} \, \Delta T_{\mathrm{ped}} \, V_{\mathrm{plasma}}\\
  \int T \, \Delta n \, \mathrm{d}V  = \beta \, T_{\mathrm{ped}} \, \Delta n_{\mathrm{ped}} \, V_{\mathrm{plasma}},
  \label{eq:integral_approx}
\end{gather}
where $\alpha$ and $\beta$ are two coefficients. Assuming $T_e = T_i$, $n_i=n_e$, $\Delta T_e = \Delta T_i$ and $\Delta n_e = \Delta n_i$, and normalizing Eq.~\ref{eq:cond_conv} to $W_{\mathrm{ped}}$, one obtains:
\begin{gather}
  \nonumber \frac{\Delta W}{W_{\mathrm{ped}}}  \approx  \frac{3 \left[\int n_e \, \Delta T_e \, \mathrm{d}V + \int T_e \, \Delta n_e \, \mathrm{d}V \right]}{3 n_{e \mathrm{, ped}} T_{e \mathrm{, ped}} V_{\mathrm{plasma}} } = \\
  = \alpha \frac{\Delta T_{e \mathrm{, ped}}}{T_{e \mathrm{, ped}}} + \beta \frac{\Delta n_{e \mathrm{, ped}}}{n_{e \mathrm{, ped}}} = \frac{\Delta W_{\mathrm{cond}}}{W_{\mathrm{ped}}} + \frac{\Delta W_{\mathrm{conv}}} {W_{\mathrm{ped}}}.
  \label{eq:approx_cond_conv}
\end{gather}
Therefore, the conductive and convective relative energy losses are related to the relative temperature and density loss at the pedestal top, respectively.
\newline Figure~\ref{Tedrop} shows the relative drop of the pedestal top electron temperature during PREs against the pedestal top collisionality for both AUG and C-Mod. The electron temperature is measured by the ECE system in both devices and 
each data point has been obtained via conditional average of several PREs in discharge phases with constant plasma parameters. A clear correlation is observed between $\Delta T_e / T_e$ and $\nu^*_{\mathrm{ped}}$, namely the relative temperature losses increase with decreasing pedestal top collisionality. For DEMO and ARC I-mode pedestal top collisionalities, the implied $\Delta T_e / T_e$ at the pedestal top ranges between 10 and 20\,$\%$. The relative drop of the pedestal top electron temperature during PREs is clearly lower than the corresponding reduction during type-I ELMs, which ranges between 20 and 50\,$\%$ across different devices~\cite{Cavedon_2017,Frassinetti_2015,Viezzer_2016, Leonard_2002, Wade_2005}. Also, it should be noted that sawtooth-triggered (squares) and non-sawtooth-triggered (diamonds) PREs do not show a particular difference in the relative electron temperature drop for similar pedestal top collisionalities.
\newline Due to Eq.~\ref{eq:approx_cond_conv}, the trend found in Fig.~\ref{Tedrop} is an indication that PRE conductive losses scale with the pedestal top collisionality, again similarly to type-I ELMs~\cite{Leonard_2002}. 
The similar correlation of both the conductive and total (conductive $+$ convective) relative energy losses with $\nu^*_{\mathrm{ped}}$ may be consistent with a weaker dependence of the PRE convective relative energy loss on $\nu^*_{\mathrm{ped}}$, as it is the case for type-I ELMs. 
\newline These findings suggest that the processes that control type-I ELM and PRE energy transport to the divertor are similar, as their total and conductive energy losses scale similarly.

\section{Energy fluence onto the divertor target}
\label{sec:enfluence}

\begin{figure}[htb]
        \centerline{\includegraphics[width=0.5 \textwidth]{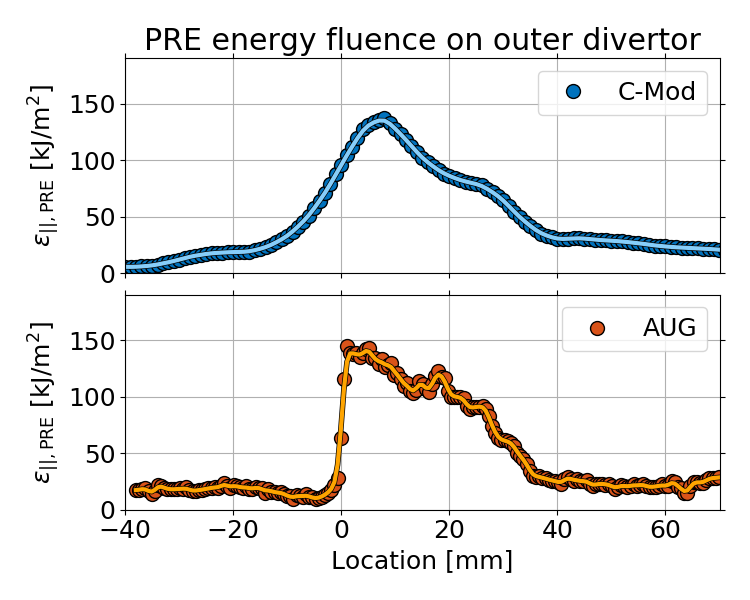}}
        \caption[]{Parallel energy fluence on the lower outer divertor target during a PRE in C-Mod (discharge $\#1120830028$ at $t=1.023$\,s) and AUG (discharge $\#37295$ at $t=4.624$\,s).}
        \label{eps_prof}
\end{figure}

\begin{figure}[htb]
        \centerline{\includegraphics[width=0.5 \textwidth]{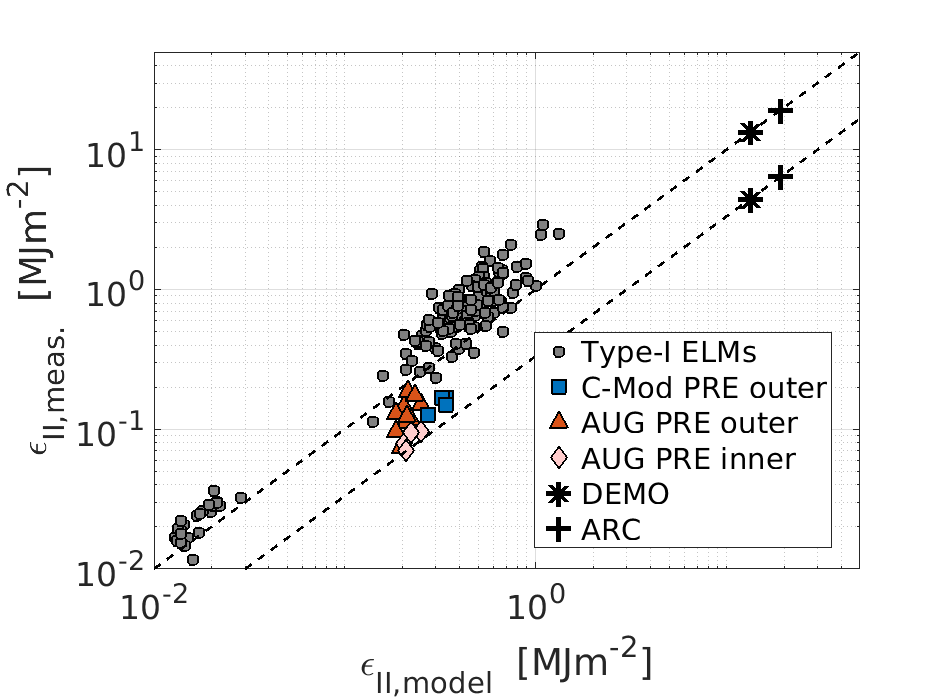}}
        \caption[]{PRE peak parallel energy fluence measured by IR cameras vs. the model prediction in AUG and C-Mod. PRE data points fall into the (1:1) and (0.33:1) lines. Stars and crosses show the prediction of the PRE peak parallel energy fluence for DEMO and ARC, respectively, using these two boundary lines. Data of type-I ELMs are shown for comparison.}
        \label{energyflu}
\end{figure}

Part of the PRE energy expelled from the confined region into the SOL ultimately reaches the divertor target plates causing a transient increase of the divertor surface temperature.
A critical parameter for the assessment of transient thermal loads on the divertor is the so-called ``heat impact factor'', which for a rectangular heat pulse is calculated as:
\begin{equation}
  \Delta T_{\mathrm{div}}  \propto  \frac{\epsilon_{\mathrm{div}}}{ \sqrt{\tau_{\mathrm{dep}}}},  
  \label{eq:impact_factor}
\end{equation}
where $\Delta T_{\mathrm{div}}$ is the divertor surface temperature rise during the transient event, $\tau_{\mathrm{dep}}$ is the deposition time and $\epsilon_{\mathrm{div}}$ is the energy fluence (J/m$^2$) deposited onto the divertor target plate. Extrapolation of these two quantities to larger devices is of crucial importance to assess the thermal load impact of I-mode PREs and to compare it to the material limit~\cite{Herrmann_2005, Yu_2016, Linke_2007}. We now investigate the divertor energy fluence due to PREs as measured on C-Mod and AUG.
\newline In both devices, the heat flux during PREs has been evaluated with infrared (IR) cameras, which measure the surface temperature of the divertor target plate. In AUG, IR cameras measure with an acquisition frequency of 1500\,Hz~\cite{Sieglin_2015}, while in C-Mod with a frequency of 380\,Hz~\cite{Terry_2010}. The space resolution of the measurement depends on which divertor target plate the camera is observing. In AUG, for the upper outer (inner) divertor tile the space resolution is 2.3\,mm/pixel (1.5\,mm/pixel), while for the lower outer divertor the resolution is 0.6\,mm/pixel~\cite{Silvagni_2020}. In C-Mod, the space resolution for the lower outer divertor is $\sim 1$\,mm/pixel~\cite{LaBombard2011}. From the measured surface temperature, the heat flux perpendicular to the divertor tile surface is evaluated with a heat diffusion equation solver. At AUG the implicit version~\cite{Nille_2018} of the THEODOR code~\cite{Sieglin_2015} is used, whereas at C-Mod the QFLUX\_2D code is implemented~\cite{LaBombard2011}. To allow cross-machine comparisons, divertor geometrical effects must be taken into account, and therefore the perpendicular heat flux is projected into the direction parallel to the magnetic field line. The parallel energy fluence is then calculated by integrating the parallel heat flux profile over the duration of the PRE:
\begin{equation}
  \epsilon_{||, \mathrm{PRE}}(s)  =  \int_{t_{\mathrm{beg}}}^{t_{\mathrm{end}}} (q_{||} - q_{||,0}) \, \mathrm{d}t.
  \label{eq:energyfluence}
\end{equation}
The inter-PRE heat flux, $q_{||,0}$, is subtracted from the parallel heat flux reaching the divertor target, $q_{||}$, to consider only the additional energy reaching the divertor due to the PRE. An example of parallel energy fluence profiles obtained with this procedure in the two devices is shown in Fig.~\ref{eps_prof}. 
\newline In C-Mod, only PREs which are not sawtooth-triggered are considered in this analysis, since the heat pulse due to the core sawtooth instability can transiently increase the energy deposited onto the divertor targets and, hence, lead to an overestimation of the PRE divertor energy fluence increase. The C-Mod database consists of PREs from 5 discharges, all of them measuring on the lower outer divertor. The AUG data in Ref.~\cite{Silvagni_NF2020} are re-examined here. They include upper divertor data from the inner and outer targets, and lower outer divertor data. The main discharge parameters of the analyzed database are summarized in table~\ref{database_fluence}.
\begin{table}[htb]
    {\footnotesize \centerline{
    \begin{tabular}{l|cc}
        \hline\hline
         &  \makecell{AUG } & \makecell{C-Mod } \\
        \hline
        $\#$ Discharges & 8 & 5 \\
        $\overline{n}_e$ ($10^{19}$ m$^{-3}$) & 3.3--5.2 & 15--17  \\
        $p_{e \mathrm{, ped}}$ (kPa) & 1.8--4.3 & 12--16 \\
        $T_{e \mathrm{, ped}}$ (keV) & 0.4--0.8 & 0.7--0.9 \\
        $q_{95}$ & 4--7 & 3 \\
        $I_{\mathrm{p}}$ (MA) & 0.6--1.0 & 1.0 \\
        $B_{\mathrm{t}}$ (T) & 2.5 & 4.3--4.8 \\
        $\delta$ & 0.2 & 0.4 \\
        $\kappa$ & 1.7 & 1.6 \\
        $R$ (m) & 1.65 & 0.65 \\
        $a$ (m) & 0.5 & 0.22 \\
        \hline\hline
    \end{tabular}}}
    \caption[]{Parameter range of the I-mode AUG and C-Mod discharges analyzed for the energy fluence database.}
    \label{database_fluence}
\end{table}
Notably, the device comparison allows variations of the major and minor radii ($R$ and $a$, respectively), of the toroidal magnetic field (and thus of $q_{95}$), and of the pedestal top parameters. In particular, the pedestal top electron pressure $p_{e\mathrm{, ped}}$ passes from $\sim 3$\,kPa in AUG to $\sim 14$\,kPa in C-Mod. Following the approach introduced in Ref.~\cite{Eich_2017}, only the peak energy fluence is considered, as this quantity needs to be directly compared to the material limits, and thus will define the allowed operational range. The experimental values of the peak energy fluence can be directly compared to those predicted by the analytical model introduced in~\cite{Eich_2017} and used to predict accurately the energy fluence found for type-I ELMs: 
 \begin{equation}
  \epsilon_{||\mathrm{, model}}  =  \Delta_{\mathrm{equi}} 2 \pi a \sqrt{\frac{1+\kappa^2}{2}} \frac{3}{2} p_{e\mathrm{, ped}} \frac{B_{\mathrm{tor}}^{\mathrm{MP}}}{B_{\mathrm{pol}}^{\mathrm{MP}}},  
  \label{eq:epsmodel}
\end{equation}
where the superscript ``MP'' stands for outer midplane and $\Delta_{\mathrm{equi}}$ is a geometrical factor derived by comparison of the assumed elliptical plasma shape to the real equilibrium reconstruction, which is $\sim $\,2.0 for AUG and $\sim $\,1.9 for C-Mod. Figure~\ref{energyflu} shows the peak parallel energy fluence measured on the divertor target during PREs against the energy fluence values given by Eq.~\ref{eq:epsmodel}. 
Type-I ELMs energy fluence values from the multi-machine study in Ref.~\cite{Eich_2017} are plotted in gray for comparison. In figure~\ref{energyflu} an additional line that shows a 0.33-times lower estimate (0.33:1) than the model predicts (1:1) is drawn.
The PRE parallel peak energy fluence is smaller than the corresponding values of type-I ELMs for the same $\epsilon_{|| \mathrm{, model}}$. Moreover, both AUG and C-Mod PRE data fall between the (1:1) and (0.33:1) lines. Even though AUG and C-Mod tokamaks have different pedestal top electron pressure values (see table~\ref{database_fluence}), the energy fluence predicted by the model in the two machines does not vary widely. This is because in C-Mod the increase in the pedestal top values is compensated by the reduction of the minor radius and of the ratio $B_{\mathrm{tor}}^{\mathrm{MP}}/B_{\mathrm{pol}}^{\mathrm{MP}}$ ($\propto q_{95}$). Therefore, the addition of C-Mod data to the AUG data does not drastically enlarge the $\epsilon_{|| \mathrm{, model}}$ variation in the analyzed database. For this reason, in order to make any conclusive statement on the scaling of the PRE peak parallel energy fluence towards a fusion power plant additional divertor energy fluence measurements are necessary. In particular, it would be envisaged to add measurements from machines with both low (or large) minor/major radius and pedestal top electron pressure values. 
\newline Nonetheless, preliminary projections to future devices could be attempted by using Eq.~\ref{eq:epsmodel} and its 0.33-times value as an upper and lower boundary, respectively. In support of this approach is not only the overall close correspondence between the PRE energy fluence measurements and the model, but also the similarity in the energy losses mechanism and scaling between type-I ELMs and PREs shown in section~\ref{sec:enloss}. 
\newline The projected peak parallel energy fluence in a DEMO I-mode scenario~\cite{Morris_2021} ranges between 4\,$-$\,13\,MJ/m$^2$, while in a ARC-relevant I-mode scenario~\cite{Sorbom_2015} it ranges between 6\,$-$\,19\,MJ/m$^2$. Those values need to be compared to the material limits, which strongly depend on the divertor target geometry design. Both ARC~\cite{Kuang_2018} and DEMO~\cite{You_2016} divertor concepts have to date been based on the ITER divertor design~\cite{Hirai_2013, Carpentier_2014}, therefore ITER material limits are used in the following discussion. The most advanced and recent divertor material limit takes into account the geometrical effect of the castellated divertor structure and predicts the limit for the perpendicular energy fluence to be $\epsilon_{\perp, \mathrm{lim}}$\,=\,0.15\,MJ/m$^{2}$~\cite{Gunn_2017}. Using an optimistic perpendicular-to-parallel conversion factor of 20, the limit parallel energy fluence at the divertor target is 3\,MJ/m$^{2}$. The lowest parallel energy fluence values predicted in DEMO and ARC are a factor 1.33 and 2 above the limit, respectively. However, regarding ARC it should be noted that the foreseen plasma configuration is a double null configuration that includes a long-leg and a secondary X-point divertor geometry in both the upper and lower divertor chambers~\cite{Kuang_2018}. It is not clear yet if I-mode can be accessed in this configuration~\cite{Hubbard_2017}, however, if yes, this configuration could introduce enhanced SOL dissipation effects and a different heat flux distribution on the targets with respect to present-day lower single null configurations. Therefore, it is expected that the above-mentioned projected values may be lowered by this divertor configuration. Also, it should be noted that the confinement scalings used to project to ARC in~\cite{Sorbom_2015} were more optimistic than those found in more recent I-mode studies~\cite{Wilks_2019}, and that projected pedestal values could exceed H-mode thresholds. Hence, the actual ARC $p_{e \mathrm{, ped}}$ and thus the energy fluence are likely to be lower.

\section{Conclusions}
\label{sec:conclusion}

In this work I-mode pedestal relaxation events (PREs) have been investigated in the Alcator C-Mod and ASDEX Upgrade tokamaks. The main results of this cross-machine comparison are the following:
\begin{enumerate}
    \item \textit{Occurrence}. I-mode PREs occur in a small subset of I-mode plasmas, i.e. in about 20\,$\%$ of the analyzed I-mode discharges. In both devices, PREs mainly appear when the plasma is close to the H-mode transition. Moreover, in C-Mod PREs can be often triggered by the heat pulse perturbation caused by core sawtooth instabilities. Notably, in C-Mod, neither I-mode PREs nor H-mode transitions were observed during discharges at $B_{t}=8$\,T in the unfavorable configuration.
    \item \textit{Triggering instability}. Before the PRE onset, a growing oscillating precursor is observed in both devices. The precursor is localized in the confined region close to the separatrix. It oscillates at a frequency roughly equal to that for the WCM in AUG, while in C-Mod the frequency of the precursor is about 50\,$\%$ larger than that of the WCM. Moreover, in C-Mod, when the plasma outer gap is reduced, the precursor is detected on the magnetic pick-up coils. These experimental findings present similarities with the simulations performed in~\cite{Manz_2021}, in which a PRE develops from the WCM that becomes interchange unstable at large plasma beta.
    \item \textit{Energy losses}. The PRE relative energy losses from the confined region increase with decreasing pedestal top collisionality $\nu_{\mathrm{ped}}^*$, similarly to type-I ELMs. The PRE energy loss is around 2--3\,$\%$ of the plasma stored energy at DEMO and ARC pedestal top collisionalities. Moreover, the relative electron temperature drop at the pedestal top -- which is related to the conductive energy losses -- exhibits a clear correlation with $\nu_{\mathrm{ped}}^*$, as type-I ELMs do. These findings suggest that the processes that control type-I ELM and PRE energy transport to the divertor are similar. 
    \item \textit{Divertor energy fluence}. The peak parallel energy fluence measured on the divertor during PREs is lower than that given by the model introduced in~\cite{Eich_2017} for type-I ELMs, which appears to represent an upper boundary of the PRE data. A lower boundary is found by dividing the model by three. These two boundaries have been used to make projections for DEMO and ARC. They give $\epsilon_{|| \mathrm{, DEMO}} = 4 - 13$\,MJ/m$^2$ and $\epsilon_{|| \mathrm{, ARC}} = 6 - 19$\,MJ/m$^2$, which are above the material limit for the ITER divertor.

\end{enumerate}

\noindent The main implication of these findings is that, if I-mode plasmas are run in a fusion power plant, then they should be run in the parameter space without PREs to avoid possible divertor thermal overloading that could lead to damage. I-modes without PREs are achieved in present-day machines by keeping the I-mode plasma away from the H-mode transition. Moreover, in a fusion power plant PREs could be used to monitor the proximity to H-mode, and hence to avoid the plasma entering an undesired ELMy H-mode.

\ack
The authors would like to thank M. Cavedon, T. Luda, P. Manz, U. Plank, F. Sciortino and M. Siccinio for fruitful discussion. This work has been carried out within the framework of the EUROfusion Consortium and has received funding from the Euratom research and training programme 2014-2018 and 2019-2020 under grant agreement No 633053. The views and opinions expressed herein do not necessarily reflect those of the European Commission. The work has also been supported by US Department of Energy, Fusion Energy Sciences, awards DE-SC0014264, DE-SC0020327, and DE-SC0014251, and that support is gratefully acknowledged.

\section*{References}

\bibliography{cit}

\end{document}